# Evolution of Insulator-Metal Phase Transitions in Epitaxial Tungsten Oxide Films during Electrolyte-Gating

*Shinichi Nishihaya, Masaki Uchida\*, Yusuke Kozuka, Yoshihiro Iwasa, and Masashi Kawasaki*

S. Nishihaya, Dr. M. Uchida, Dr. Y. Kozuka, Prof. Y. Iwasa, and Prof. M. Kawasaki

Department of Applied Physics and Quantum-Phase Electronics Center (QPEC), University of Tokyo, Tokyo, 113-8656, Japan

Prof. Y. Iwasa, and Prof. M. Kawasaki

RIKEN Center for Emergent Matter Science (CEMS), Wako, 351-0198, Japan

ABSTRACT

An interface between an oxide and an electrolyte gives rise to various processes as exemplified by electrostatic charge accumulation/depletion and electrochemical reactions such as intercalation/decalation under electric field. Here we directly compare typical device operations of those in electric double layer transistor geometry by adopting *A*-site vacant perovskite $WO_3$ epitaxial thin films as a channel material and two different electrolytes as gating agent. *In situ* measurements of x-ray diffraction and channel resistance performed during the gating revealed that in both the cases $WO_3$ thin film reaches a new metallic state through multiple phase transitions, accompanied by the change in out-of-plane lattice constant. Electrons are electrostatically accumulated from the interface side with an ionic liquid, while alkaline metal ions are more uniformly intercalated into the film with a polymer electrolyte. We systematically demonstrate this difference in the electrostatic and electrochemical processes, by comparing doped carrier density, lattice deformation behavior, and time constant of the phase transitions.

1. INTRODUCTION

Interfaces between solids and electrolytes have shown various types of chemical phenomena such as crystal growth,[1] catalytic process,[2] and redox reaction[3] through complex exchange of electrons, ions, and energy. Especially under electric fields, ions can be incorporated into host materials, which is known to be a key reaction in ion battery.[4] Besides such an energy storage application, intercalation has been also established as an effective way to control physical properties and electronic phases of host compounds. For example, phase transitions into such exotic electronic phases as a superconducting state have been induced by intercalation in some porous or layered compounds.[5-8]

In contrast to such chemical reactions, there is also a case that ions or charged molecules are accumulated on the solid surface to form so-called electric double layer (EDL) at the interface.[9] This gives rise to electrostatic accumulation of charge carriers on the solid surface, known as the operation mechanism of the EDL transistor (EDLT). EDLT takes advantage of the subnanometer-gap capacitor within EDL, resulting in exceptional capability of charge carrier accumulation up to a sheet carrier density of $10^{15}$ cm$^{-2}$ on the channel surface.[9] It has been reported so far that superconductivity can be induced by the field effect in band insulator oxides such as $SrTiO_3$[10] and $KTaO_3$.[11] However, large electric field within EDL could also induce some chemical reactions such as intercalation of ions from the electrolyte or removal of ions from the solid surface. There are many unsolved questions on the origin of phase transitions at the interface of transistors employing electrolyte-gating, while they have been extensively examined in various oxides.[12-14]

In this context, it is highly meaningful to pick up a channel solid applicable for both the electrostatic accumulation and electrochemical intercalation experiments in order to directly compare the gating processes. For this purpose, we choose a simple band insulator oxide $WO_3$. While the carrier doping in $WO_3$ has attracted attentions mainly in terms of doping induced electrochromism,[15] here we focus on the point that $WO_3$ has a perovskite type structure with vacant *A*-sites. Due to this remarkable feature, abundant vacancies in the lattice can accomodate cations as in the anode of ion battery.[16,17] By using electrolyte containing alkaline metal ions such as Li$^+$, ion intercalation can be easily induced by applying a gate voltage. Therefore, $WO_3$ is suitable for comparing the doping processes between the electrostatic carrier accumulation and the electrochemical intercalation. Moreover, it has been well known that $WO_3$ easily changes its lattice structure depending on the density of carriers or dopants.[18-20] This enables us to investigate the two gating processes not only by electrical transport properties but also in terms of the lattice deformation.

Here we study the electric field gating of $WO_3$ thin film with using two different types of electrolytes as gating agent in EDLT. One is a typical ionic liquid, N,N-diethyl-N-(2-methoxyethyl)-N-methylammonium bis-trifluoromethylsulfonyl-imide (DEME-TFSI) that is expected to accumulate electrons purely by electrostatic process. The other is a polyethylene oxid (PEO) containing $LiClO_4$ as a supporting salt, where Li ions are expected to be intercalated into the $WO_3$ thin film to dope electrons. Clear insulator-metal transitions are observed in both the cases. We have also performed an *in situ* x-ray diffraction (XRD) measurement to investigate the lattice deformation during the two gating processes. Since the in-plane lattice of fabricated $WO_3$ thin film is epitaxially strained to match coherently with that of the substrate, any structural deformation along the in-plane directions is forbidden. Under this condition, it is revealed that a couple of new phases with different *c*-axis lengths successively appear depending on the gate voltage. The observed phase transition behaviors between these phases clearly reflect the difference in the two gating processes.

2. RESULTS AND DISCUSSION

$WO_3$ thin films were grown on $YAlO_3$ substrates by pulsed laser deposition, yielding in epitaxial films as summarized in **Fiugre 1**. From the XRD $2\theta$-$\omega$ scan, the film is confirmed to have a *c*-axis oriented single phase (**Figure 1a**). Laue fringes are also observed around each diffraction peak. The FWHM of (001) rocking curve is as narrow as 0.02 degrees, ensuring excellent crystallinity (**Figure 1b**). As shown in reciprocal space mapping (**Figure 1c**), the film is coherently grown with in-plane *a*- and *b*-axes tensely strained by 0.4 %, while the out-of-plane *c*-axis is elastically compressed by 1.0 % compared to the bulk one. In this situation, only the *c*-axis lattice constant changes through the carrier doping. **Figure 1d** shows the surface morphology, taken by an atomic force microscopy (AFM), indicating 0.4 nm steps and atomically flat terraces (root mean square roughness of about 0.2 nm). **Figure 1e** shows a reflection high energy electron diffraction (RHEED) pattern of the film just after the growth. Fine spots on first Laue zone and Kikuchi lines are clearly obeserved, also indicating the high crystalline perfection.

    We have fabricated electrolyte-gating devices using the $WO_3$ thin films. DEME-TFSI is used as a gate material for electrostatic charge accumulation whereas a PEO electrolyte containing $LiClO_4$ for Li intercalation. In the following results and discussions, we refer the case of DEME-TFSI as "ionic liquid", and the case of PEO+$LiClO_4$ as "polymer electrolyte". **Figures 2a and 2b** show a typical Hall-bar structure fabricated for the transport measurement. We have also prepared a different type of device for *in situ* XRD measurement, as shown in **Figures 2c and 2d**. **Figures 2e and 2f** illustrate schematic models of the doping processes in ionic liquid and polymer electrolyte experiments. For the former, an electric double layer is formed in EDLT and electrons are accumulated on the film surface when the gate is positively biased. For the latter, on the other hand, Li ions are expected to be incorporated into the vacant *A*-sites in $WO_3$ lattice.

    **Figures 3a (3b)** shows transfer curves during charging and discharging of electrons (Li ions) in the ionic liquid (polymer electrolyte) device. In those experiments, the drain source current $I_D$ shows typical n-type characteristics and increases by more than 3 orders of magnitude as positively increasing the gate voltage $V_G$. The leakage current $I_G$ during the gating process is

much larger in polymer electrolyte device than in ionic liquid one, which respectively reflects electrochemical current due to intercalation of Li ions and displacement current in electrostatic accumulation. **Figures 3c and 3d** show the temperature dependence of the resistivity down to 2 K. As $V_G$ is increased, the resistivity decreases and a transition into a metallic state is observed for both the cases. The critical value of $V_G$ is about 2.0 V in ionic liquid and 1.6 V in polymer electrolyte experiments. Around the critical value, the change of resistivity by carrier doping tends to be small compared to that in the insulating region, which has also been observed in the case of chemically doped $WO_3$.[21,22] The critical $V_G$ for metallization of $WO_3$ in the ionic liquid device actually changes depending on the surface area ratio of $WO_3$ channel and gate electrode as shown in the following *in situ* XRD measurements and this tendency clearly indicates electric double layers formation at the interfaces. From the Hall measurements at 2 K (**Figures 3e and 3f**), the carrier type is confirmed as electrons and the density is estimated to be as larger as $3.4 \times 10^{22}$ cm$^{-3}$ at maximum in ionic liquid device, even assuming that the films are uniformly doped. This carrier density is surprisingly large, reaching almost 2 electrons per unit cell. Compared with other report on ionic liquid gating of hexagonal $WO_3$ films,[23] the carrier density is one order of magnitude higher in our study. The origin of this high carrier density is discussed later. In contrast, the estimated carrier density for polymer electrolyte device remains less than 1 electron per unit cell at maximum. This value is reasonable from the aspect of intercalation process. The maximum carrier density by intercalation should be physically limited to 1 per unit cell because the vacancy of $WO_3$ lattice (i.e. *A*-site of $ABO_3$ perovskite lattice) can only accommodate Li ions up to 1 per unit cell. We note that a large hysteresis is also observed in the transfer curve in the ionic liquid case, which may seem to contradict with an ideal picture of electrostatic charge accumulation. As for the origin of the hysteresis, we point out that the gate-induced metal-insulator transitions in $WO_3$ films are the first-order transitions accompanying structural changes as shown in later XRD experiments. There should be energy barriers between insulating and metallic phases and thus it is natural that $V_G$ or carrier density necessary for the transitions between these phases becomes different depending on sweep directions and gives rise to such a hysteresis in the transfer curve. Any adsorption of ionic liquid molecules on the channel surface to give the persistent conducting state at $V_G = 0$ V may be also possible, as reported in ionic liquid gating for $VO_2$ thin film.[24]

To investigate the charging and discharging effects on the crystal structure, we have performed *in situ* XRD measurement for ionic liquid and polymer electrolyte devices using the configurations shown in **Figures 2c and 2d**. **Figures 4a and 4b** summarize $2\theta$-$\omega$ scans around the $WO_3$ (002) peak, showing distinct structural changes between both devices. First, $V_G$ is increased stepwise up to 4.0 V for the ionic liquid device and to 1.8 V for the polymer electrolyte one for charging and then $V_G$ is switched to 0 V (short circuit) to investigate the time dependence for discharging. In order to analyze the results, *c*-axis/*a*-axis ratio (*c*/*a* ratio) and area of the diffraction peaks were calculated by fitting the diffraction peaks with Gaussian. The results for the whole gating processes including $V_G$, resistance, *c*/*a* ratio and peak area are plotted against the time in **Figure 5a** for the ionic liquid and **5b** for the polymer electrolyte devices.

First we focus on the results of ionic liquid experiments. As $V_G$ is increased in charging process, there appear three phases in total, namely, phase 1 with *c*/*a* ~ 1.035, phase 2 with ~ 1.045 and phase 3 with ~ 1.022 across first order transitions. Compared with phase 1, *c*-axis extends in phase 2 and shrinks in phase 3 by about 1%. These phases coexist upon phase transitions from phase 1 to 2 and also from phase 2 to 3 at around $V_G = 1.5$ V and 3.0 V respectively. These abrupt changes in *c*-axis length accompanied by the phase separation are reproduced in the discharging process. In a bulk $WO_3$ crystal with monoclinic symmetry, mainly *a*- and *b*-axes are elongated monotonically towards highly symmetric lattice with increasing the doping level.[18,19] In the present case, on the other hand, the in-plane lattice of the film is coherently clamped to the substrate, which is also confirmed during the gating process (not shown). Thus, the forbidden in-plane lattice deformation is speculated to result in the non-monotonic lattice constant change in *c*-axis direction. We note that the shrinkage of *c*-axis in highly doped phase 3 is consistent with the bulk case in the point that the lattice is approaching cubic symmetry ($c/a = 1$).

Next we move on to the results of polymer electrolyte experiments and compare them with those of ionic liquid ones. Extension and shrinkage of the *c*-axis are also observed as shown in **Figures 4b and 5b**. However, the phase transition processes are quite different from those in the ionic liquid experiments. When the *c*-axis is elongated, the system shows continuous change of the *c*-axis in polymer electrolyte device rather than the phase separation into phases 1 and 2 in ionic liquid case. We refer this region as phase1'. This difference is indeed a clear evidence of the different doping processes in ionic liquid and polymer electrolyte experiments. EDLs formed on the channel surface of the ionic liquid device result in the lattice change starting from the electron accumulated layer, and then the phase transition propagates into the bulk as the carrier density is increased. For the polymer electrolyte device, in contrast, Li ions diffuse through the whole film, which leads to the uniform change in lattice structure of the whole film. Thus, the difference in the transitons is ascribable to the difference between electrostatic carrier accumulation and electrochemical intercalation. As for the *c*-axis shrunk state, on the other hand, intercalation shows the same phase separation as electrostatic accumulation. Let us refer the *c*-axis shrunk state as phase 3' for polymer electrolyte experiments.

In addition to this difference in *c*-axis elongation, we also find a clear difference in discharging processes between those two devices. For the ionic liquid device, the film returns back to the original insulating state about 200 min after $V_G$ is turned to 0 V. For the polymer electrolyte device, on the other hand, the resistance remains low even for 1000 min after the turn-off. Only after negative bias of $V_G = -0.5$ V is applied, considerable increase of resistance is observed. This suggests that most of the intercalated Li ions remain in the $WO_3$ lattice and phase 3' keeps dominant even at $V_G = 0$ V, while they are quickly decalated under the negative bias of $V_G = -0.5$ V.

Common behavior obeserved for *c*-axis shrunk phase 3 in ionic liquid case and phase 3'in polymer electrolyte case implies that they have a same origin, although their electron doping processes are different. These *c*-axis shrunk states tend to appear only after the resistance change slows down against the increase of $V_G$. According to the resistivity measurements shown in **Figures 3c and 3d**, the resistivity saturated points against $V_G$ are near the insulator-metal transition. From this comparison, the *c*-axis shrinkage is considered to be induced by the electronic phase transition, namely insulator-metal transition in $WO_3$ film. To clarify this point, we have also measured the temperature dependence of the resistance in each phase. As exemplified in **Figure 6**, insulator-metal transition clearly occurs together with the appearance of phase 3 in ionic liquid device. Therefore, the *c*-axis shrunk state is indeed identified as a metallic state, which is realized for the first time in epitaxially strained $WO_3$ film. We thus conclude that the first transition (phase 1 to 2 in ionic liquid device and within phase 1' in polymer electrolyte device) results from the volume changes due to electron doping, reflecting the difference in doping processes, while the second one (phase 2 to 3 in ionic liquid and phase 1' to 3' in polymer electrolyte) corresponds to the lattice deformation driven by electronic phase transition, namely insulator-metal transition which occurs independently on the way of electon doping.

It is also notable that the $V_G$ necessary to achieve metallization or same carrier density is almost doubled compared to the results of transport measurement of the ionic liquid device shown in **Figure 3**. Assuming an electrostatic process, this change in critical $V_G$ can be ascribed to different surface area ratio between WO$_3$ channel and gate electrode ($A_{WO3}/A_{ele}$) in each device configuration. In the Hall bar device used for transport measurements shown in **Figure 3**, $A_{WO3}/A_{ele}$ is almost 0 (0.03 mm$^2$/ 60 mm$^2$), thus the total capacitance of the system is dominated by that of WO$_3$/electrolyte interface. Then $V_G$ is effectively applied to the WO$_3$/electrolyte interface to accumulate charges. On the other hand, $A_{WO3}/A_{ele}$ is 1 (6 mm$^2$/ 6 mm$^2$) in *in situ* XRD device. This leads to the comparable contributions to the total capacitance from electrode/electrolyte and WO$_3$/electrolyte interfaces, reducing the effective voltage at the latter interface by half. Therefore, the difference of effective $V_G$ at WO$_3$/electrolyte interface due to $A_{WO3}/A_{ele}$ in each device also implies the electrostatic process in the ionic liquid device.

From the point of view of electrostatic electron accumulation, one unsolved question would be the origin of such a high sheet carrier density as $2.3 \times 10^{16}$ cm$^{-2}$ estimated by the Hall measurements, since the accumulation of the same amount of counter cations at the other side of the interface is electrostatically unfeasible. As one possibility, the value $1/eR_H$, where $R_H$ is Hall coefficient and $e$ is elementary charge, may diverge from simple carrier density and the real density is overestimated in the heavily doped region. This can happen in some cases. One is when compensation of electrons and holes occurs in semimetals such as graphene.[25] This is, however, clearly not the case for our results, since appearance of holes as second carrier is unlikely in the gating process for electron accumulation. The divergence of $1/eR_H$ from simple carrier density is also reported for 2 dimensional electron gas at LaAlO$_3$/SrTiO$_3$ interface[26] and it is attributed to the criticality on the verge of Mott transition of (La,Sr)TiO$_3$.[27] The high accumulation of electrons may enhance electron correlation also in the 5$d$ orbitals of WO$_3$, but it is not clear whether or not such Mott insulator phase emerge in heavily doped WO$_3$. Further investigations are necessary to clarify the origin of the large 1/$eR_H$ value.

Nonetheless, our results in the transport and the *in situ* XRD measurements clearly capture the characteristic aspects of electrostatic carrier accumulation in the ionic liquid device as well as those of electrochemical intercalation in polymer electrolyte device. Success in a direct comparison of the two doping processes by utilizing WO$_3$ as channel material can lead to more profound understanding of the underlying dynamics at the oxide/electrolyte interface.

## 3. CONCLUSION

In summary, we have systematically compared the device operations with electrostatic accumulation/depletion and electrochemical intercalation/decalation processes, by adopting *A*-site vacant perovskite WO$_3$ film as a channel material and two different electrolytes as gating agent. In both the cases, insulator-metal transition has been observed when the gate voltage is above a critical value. As the film is doped with electrons, the *c*-axis non-monotonically extends and shrinks, reflecting the strong correlation between the lattice structure and doping level in WO$_3$. A clear difference appears in the beginning of the two doping processes; the carrier-accumulated layers spread from the interface to bring the phase separeation in the former, while intercalated Li ions diffuse through the whole film and induce the continuous phase change in the latter. This difference is also clearly indicated in the depletion and decalation processes, where the former occurs just setting $V_G$ = 0 V and the latter requires a driving force to decalate. We have successfully distinguished the electrostatic and electrochemical processes in terms of the electrical and structural phase changes. Our findings and techniques will pave the way for a comprehensive understanding of complex doping dynamics at the interface between oxide films and electrolytes.

4. METHODS

WO$_3$ thin films were grown on (110) surfaces of single-crystal YAlO$_3$ substrates by pulsed laser deposition. The laser fluence and repetition frequency were fixed at 0.35 Jcm$^{-2}$ and 1 Hz. The substrate temperature and oxygen partial pressure were kept at 600°C and 100 mTorr during the deposition. An ionic liquid, N, N-diethyl-N-(2-methoxyethyl)-N-methylammonium bis-trifluoromethylsulfonyl-imide (DEME-TFSI) (Kanto Chemical Co.) was used for electrostatic experiments. The polymer electrolyte for electrochemical experiments was prepared by dissolving LiClO$_4$ (Wako Pure Chemical Industries Ltd.) into polyethylene oxide with a mean molecular weight of 600 (Sigma-Aldrich Co.), yielding in the molar ratio of Li to O of 1: 20. For dehydration, the electrolytes were kept over 70°C in vacuum for several days before use.

For transport measurement, WO$_3$ films were processed into devices with a Hall-bar shape by photolithography and Ar ion milling. The film thickness was 6.7 nm for the ionic liquid and 4.9 nm for the polymer electrolyte devices, and the typical channel size was 60×30 μm$^2$. Ti (100 nm)/Au (500 nm) metal electrodes were then evaporated for electrical contacts. In addition, a separator, which is a hard baked photoresist layer (2 μm) for the ionic liquid device and a SiO$_2$ (300 nm)/Ti (100 nm) layer for the polymer electrolyte device, was deposited on the metal pads to prevent damages from unexpected chemical reactions with the electrolytes. A Pt coil was set as a gate electrode above the channel (**Figures 2a and 2b**).

Transport properties were measured using an Agilent 4155C semiconductor parameter analyzer. The Hall-bar device was kept and cooled in a Physical Property Measurement System (PPMS, Quantum Design) in helium gas of 1 Torr. Transfer curves were taken at 300 K, with sweeping $V_G$ between ±2.0 V at a sweep rate of 10 mVs$^{-1}$. When measuring temperature dependence of the resistivity down to 2 K, drain source voltage $V_D$ was kept at 0.5 V and $V_G$ was applied at 300 K for typically 30 min until $I_D$ was saturated. The Hall resistivity was also measured at $V_D = 0.5$ V.

*In situ* XRD measurement was performed at room temperature. The device was set in a home-made vacuum cell with a highly x-ray transparent plastic window. Inside of the cell was evacuated below 3 Torr. Several lead wires come out from the cell, enabling us to make electrical contacts to the device from the outside. To leave the x-ray path open on the channel area, the gate electrode was evaporated aside on the film-free area of the substrate (**Figures 2c and 2d**). The film thickness was 18 nm and the channel size was 2×3 mm$^2$. $V_G$ was increased at 0.25 V step, after $I_D$ was confirmed to be saturated at $V_D = 0.1$ V. In the comparison experiment between phase changes in *in situ* XRD and temperature denpendence of the resistance shown in **Figure 6**, we used the samples cut from the same film with a film thickness of 13nm and adopted the *in situ* XRD device configuration for both the measurements.

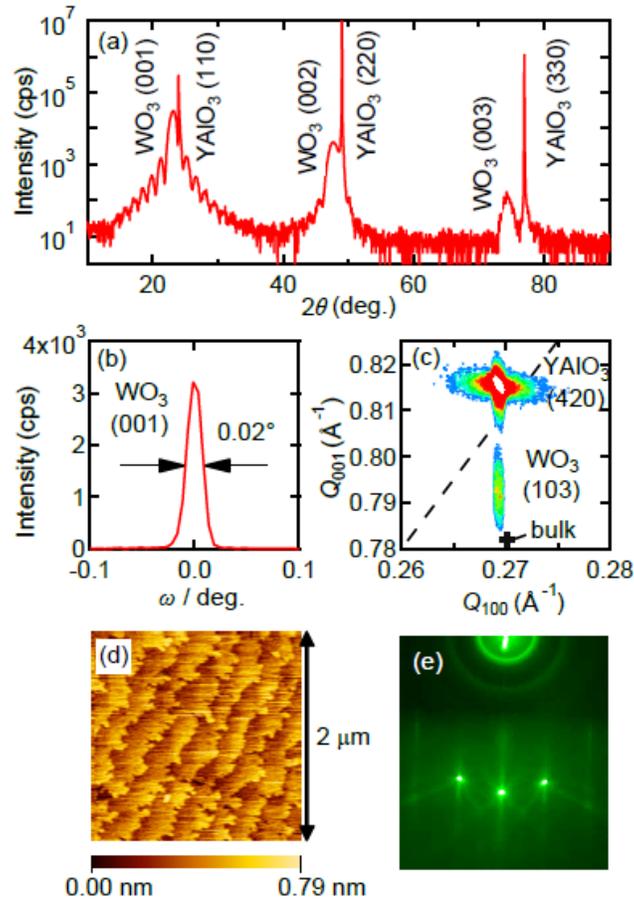

**Figure 1.** Structural characterization of $WO_3$ thin film grown on $YAlO_3$ substrate. (a) $2\theta$-$\omega$ scan of XRD. (b) Rocking curve around $WO_3$ (001) peak. (c) Reciprocal space mapping around $WO_3$ (103) peak. The broken line denotes cubic lattice symmetry. (d) AFM image of the film surface. (e) RHEED pattern of the film taken with [001] incidence.

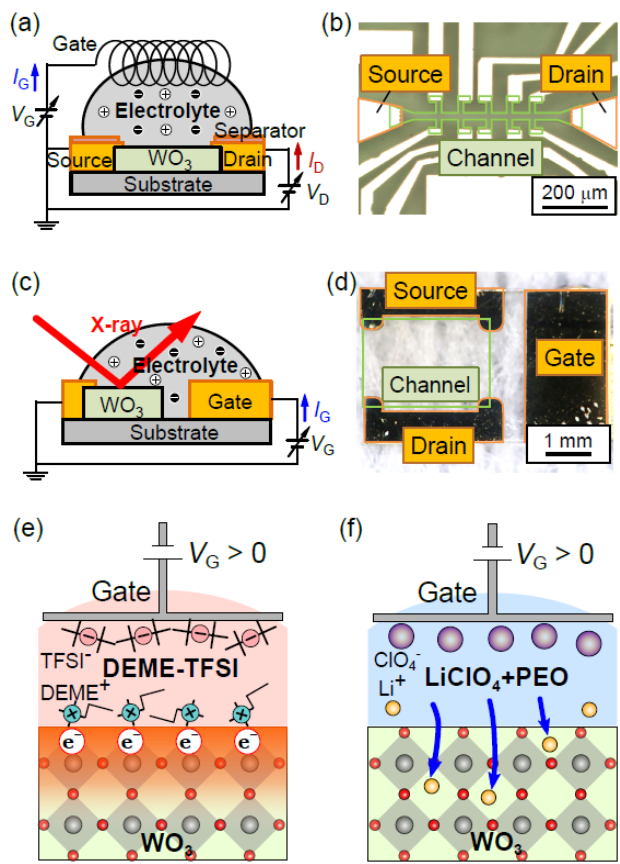

**Figure 2.** Device configuration and schematics of two doping processes. (a) Cross-sectional schematic and (b) top view picture of a Hall bar device. The channel size between the adjacent terminals is 60×30 μm$^2$. (c) Cross-sectional schematic and (d) top view picture of a device fabricated for *in situ* XRD measurement. The channel size is 2×3 mm$^2$. Schematics of (e) electrostatic accumulation of electrons with an ionic liquid and (f) intercalation of Li ions into the vacant *A*-sites of WO$_3$ lattice from a supporting salt in a polymer electrolyte.
15

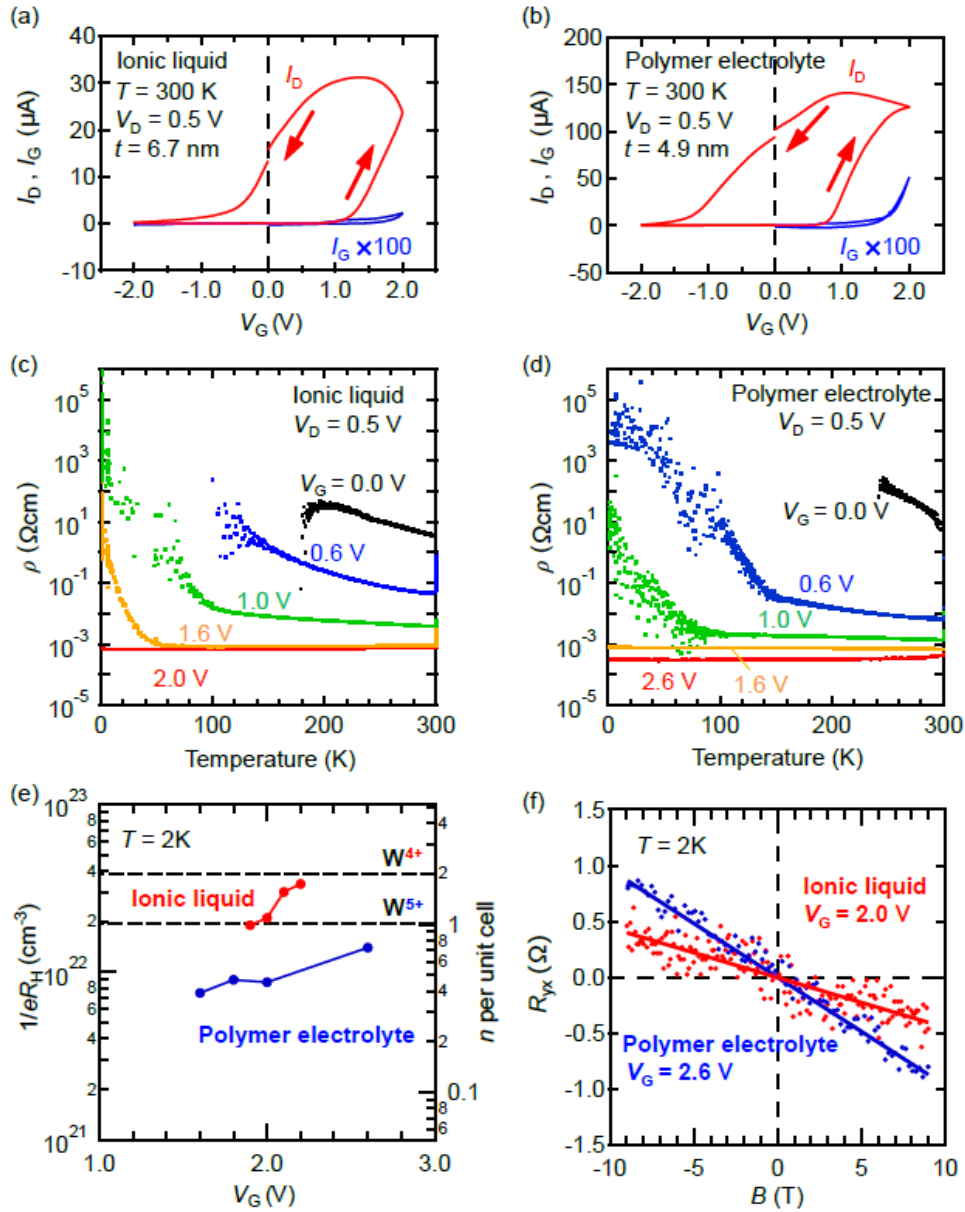

**Figure 3.** Transport measurement for the Hall-bar devices. Transfer behavior of the transistors for (a) ionic liquid and (b) polymer electrolyte devices at 300 K. Temperature dependence of the resistivity for (c) ionic liquid and (d) polymer electrolyte devices. (e) Carrier density in metallic region, derived from the Hall measurements at 2 K. The corresponding number of electrons per unit cell is estimated on the right axis. (f) Hall resistance vs magnetic field at $V_G = 2.2$ V for the ionic liquid and at 2.6 V for the polymer electrolyte devices.

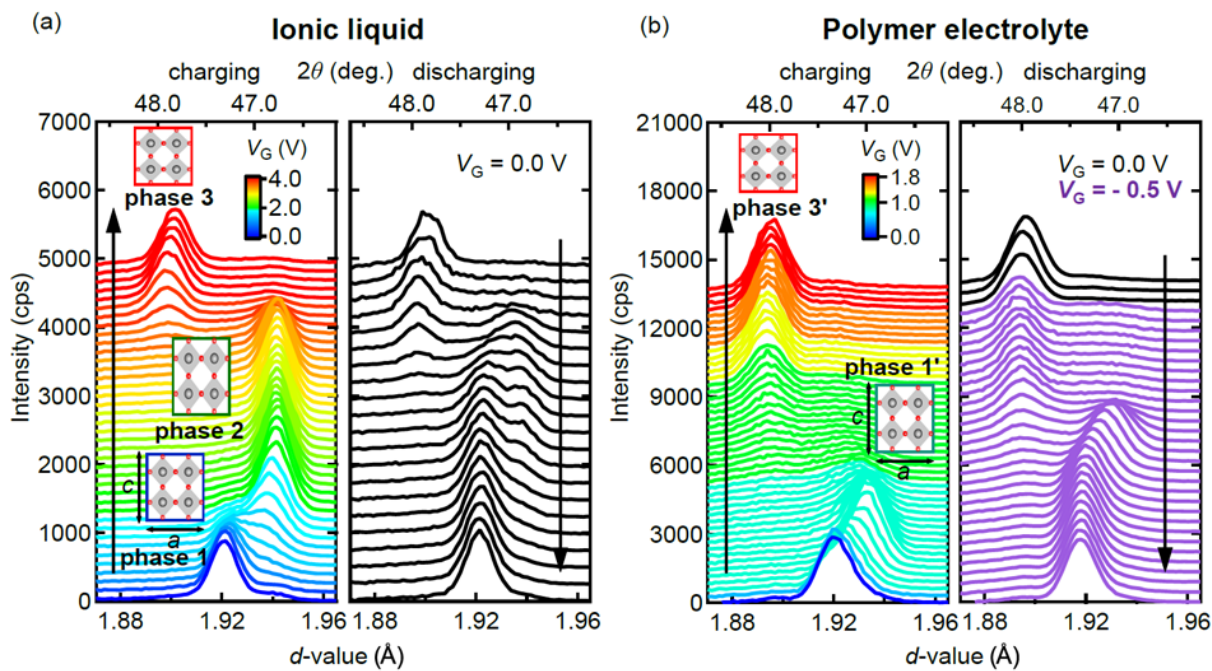

**Figure 4.** *In situ* XRD spectra during (a) electrostatic accumulation/depletion for the ionic liquid device and (b) intercalation/decalation for the polymer electrolyte device. The left panels are for charging process increasing $V_G$ stepwise. The right panels are for discharging process after switching the $V_G$ from on-state ($V_G > 0$ V) to off-state ($V_G = 0$ V). For the polymer electrolyte device, $V_G$ is decreased to −0.5 V in order to complete decalation. Background intensity from the substrate is subtracted from the raw data. The arrows denote the stream of time. The data are shifted vertically for clarity.

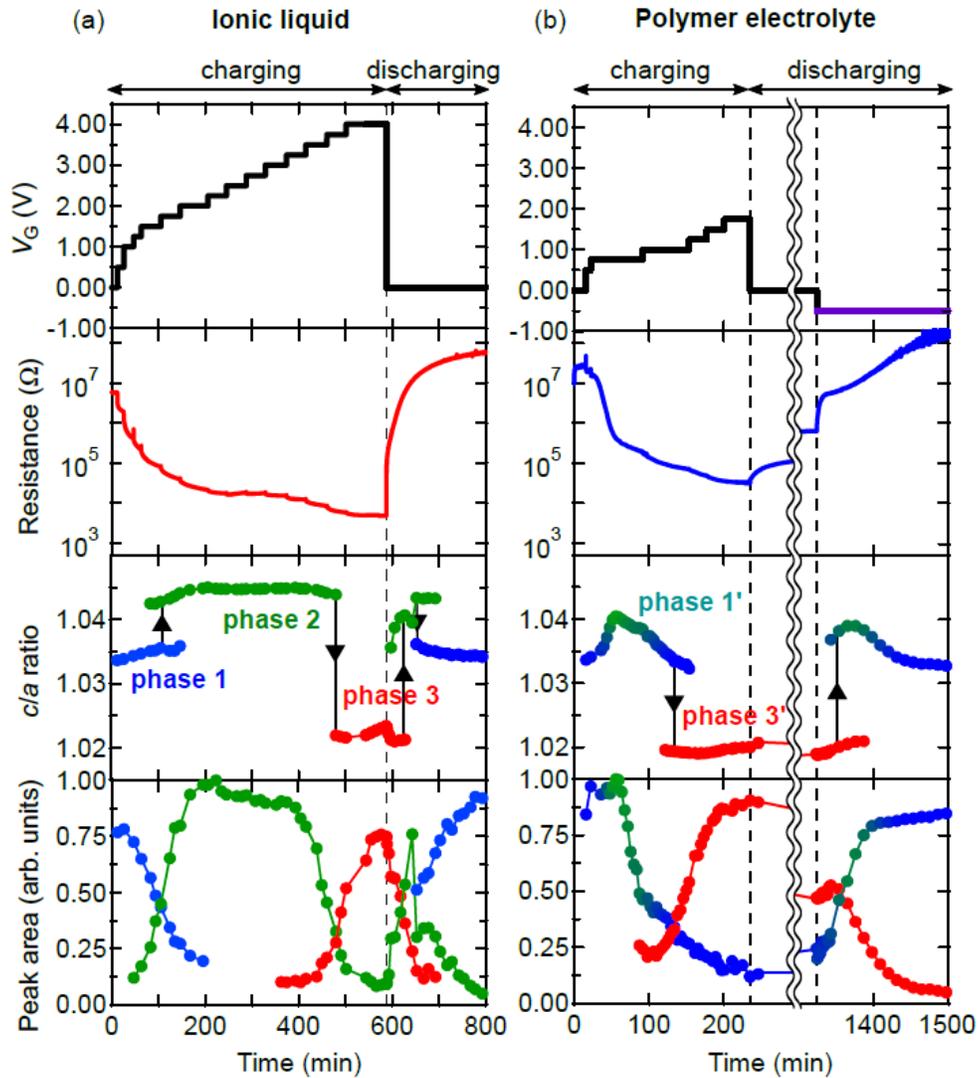

**Figure 5.** Temporal variation of the gate voltage $V_G$, resistance, $c/a$ ratio, and peak area of each phase during the whole gating processes for (a) ionic liquid and (b) polymer electrolyte. The peak area is normalized by the maximum value through each measurement. For clarity, the data points with peak area over 0.25 are shown in the plots of $c/a$ ratio vs time. The data points for phase 1 are shown in blue, phase 2 in green and phase 3 in red for ionic liquid device, while for the polymer electrolyte device the color changes continuously from blue to green within phase 1'.

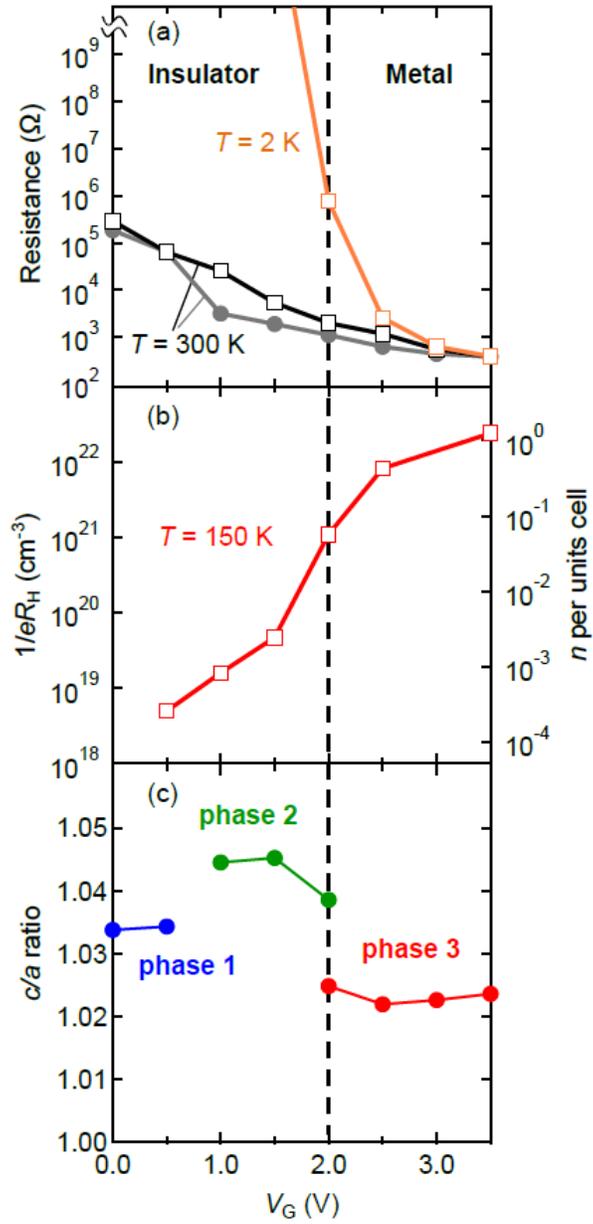

**Figure 6.** $V_G$ dependence of (a) resistance, (b) carrier density, and (c) $c/a$ ratio for the ionic liquid device. Data points in the temperature dependence measurement are represented by open squares, while those in the *in situ* XRD measurement by solid circles.

AUTHOR INFORMATION

**Corresponding Author**

*Masaki Uchida E-mail: uchida@ap.t.u-tokyo.ac.jp

**Author Contributions**

S.N. and M.U. performed the experiments, analyzed the data, and wrote the manuscript. S.N., M.U., Y.K., Y.I. and M.K. discussed results and manuscript. M.K. supervised the project. All authors have approved the final version of the manuscript.

ACKNOWLEDGMENT

This work was partly supported by Grant-in-Aids for Specially Promoted Research No. 25000003, Scientific Research (S) No. 24226002, Young Scientists (A) No. 15H05425, and Challenging Exploratory Research No. 26610098 from MEXT, Japan, and by the Murata Science Foundation and the Asahi Glass Foundation.

Table Of Contents graphic;

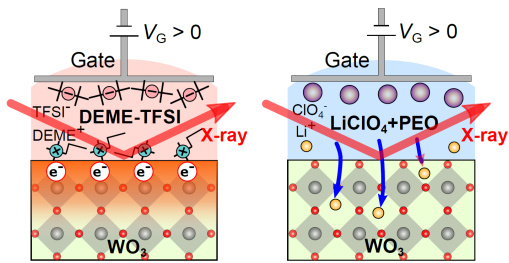